\documentclass[epsfig,psfig,rgb]{aa}

\usepackage{graphicx}
\usepackage{txfonts}
\begin{document}

\title{An empirical calibration of Lick indices using Milky Way Globular Clusters\thanks{Based on observations
collected at the European Southern Observatory, Chile, programs 0.77B0195(A) and (B).}}
\titlerunning{Calibration of Lick indices in GCs}

\author{A.Pipino$^{1,3}$, I.J. Danziger$^{1,2}$}
\authorrunning{Pipino \& Danziger}

\institute{
$^1$Dipartimento di Astronomia, Universita di Trieste,
    Via G.B. Tiepolo, 11, I-34127, Trieste, Italy \\
$^2$INAF- Osservatorio Astronomico di Trieste,Via G.B. Tiepolo, 11, I-34127, Trieste, Italy\\
$^3$Institut fur Astronomie, ETH Zurich, Wolfgang-Pauli-Str. 37, 8092 Zurich, Switzerland}
\date{Accepted,
      Received }


\abstract{}{To provide an empirical calibration relation in order to convert Lick
indices into abundances for the integrated light of old, simple
stellar populations for a large range in the \emph{observed} [Fe/H] and [$\alpha$/Fe]. 
This calibration supersedes the previously adopted ones because it is be based
on the real abundance pattern of the stars instead of the commonly adopted
metallicity scale derived from the colours.}{We carried out a long-slit spectroscopic study of 23
Galactic globular cluster for which detailed chemical abundances
in stars have been recently measured. The line-strength indices, as coded
by the \emph{Lick system} and by Serven et al. (2005), were measured in low-resolution integrated spectra of the GC light.
The results were compared to average abundances and abundance ratios in stars taken from the compilation
by Pritzl et al. (2005) as well as to synthetic models.}
{Fe-related indices grow linearly as a function of [Fe/H] for [Fe/H]$>-$2. Mg-related indices
respond in a similar way to [Mg/H] variations, however Mgb turns out to be a less reliable
metallicity indicator for [Z/H]$<-$1.5\, .
Despite the known Mg overabundance with respect to Fe in GC stars, it proved impossible
to infer a mean [Mg/Fe] for integrated spectra that correlates with the resolved stars properties, because
the sensitivity of the indices to [Mg/Fe] is smaller at lower metallicities. We present empirical
calibrations for Ca, TiO, Ba and Eu indices as well as the measurements of $H_{\alpha}$ and NaD.}{}

\maketitle

\keywords{Galaxy: globular clusters: general, Galaxy: globular clusters: individual: NGC104, NGC362,
NGC3201, M68, NGC4833, M5, M80, M4, M12, M10, NGC6287, NGC6293, NGC6342, NGC6352, NGC6362, NGC6397,
NGC6528, NGC6541, NGC6553, M22, M54, NGC6752, M30, Galaxies: elliptical and lenticular, cD}

\section{Introduction}

Simple Stellar Populations (SSPs), namely  
populations of stars characterized by the same Initial Mass Function (IMF), 
age and chemical composition, are the building blocks of both real and model galaxies,
which are made of a mixture of several SSPs, 
differing in age and chemical composition according to the galactic chemical enrichment history,
weighted with the star formation rate.
Therefore, unraveling the information encoded in the SSPs provides direct insights into 
different aspects of the galactic evolution.
Unfortunately, for most of the galaxies we cannot resolve single
stars. The only available tool is to observe the integrated light
coming from all the stars and try to infer knowledge of the star
formation history and of the chemical enrichment by means of either their colours
or their spectra. This kind of diagnostic suffers from the well-known
age-metallicity degeneracy.
A useful tool was suggested by the Lick group with a system of line-strength indices (Worthey et al.,
1994). It provides a set of 25 indices which
help in disentangling the age effects from the metallicity ones.
The problem is, however, further complicated by the fact that a third
parameter, namely the [$\alpha$/Fe] abundance ratio, gives a
non negligible contribution to the line indices.

Only recently
theoretical line-strength indices tabulated for SSPs as functions of their age, 
metallicity and $\alpha$-enhancement have been published (Thomas et al., 2003, Schiavon, 2007, Lee et
al. 2007, Thomas et al., 2010).

Globular clusters (GCs) are probably the closest approximation to a Single Stellar Population.
Lick indices have been measured in metal poor GCs (Burstein et al., 1984, Covino
et al., 1995, Trager et al., 1998) and Galactic Bulge metal rich GCs (Puzia et al., 2002), with
a recent large homogeneous sample being the one by Schiavon et al. (2005).
Synthetic SSPs have usually been calibrated on the Galactic GCs (e.g. Maraston et al. 2003,
Lee et al., 2009). In many cases, however, the adopted \emph{intrinsic} metallicity scale was based on 
some generic metallicity labelled as ``[Fe/H]'' and inferred mostly from photometric measurements (e.g. Zinn \& West, 1984, see also Harris, 1996).

Now that high-resolution spectra of single stars in GCs
are available, it is possible to test the accuracy of the SSP prediction
as well as the reliability of the ``inversion''(namely from measured indices
to inferred abundances) technique against the \emph{true} [Fe/H] (or [Mg/Fe]).
It turns out that while the same SSPs that are assumed to be calibrated on the Galactic
GCs perform quite well in recovering the metallicity, they fail in recovering the
abundance ratios.
For instance, by looking at the results by Mendel et al. (2007, their fig. 5),
who tested the inversion technique for a number of widely used SSPs against the observed
abundances in the stars of the GCs, one notes that neither any of the adopted stellar population models (including the most widely
used by the community) correlate with the [Mg/Fe] observed in the stars of the globular clusters,
nor do they reproduce the [Mg/Fe]-[Fe/H] relation (their fig. 6) which is clear also for the MW globular clusters. 
The only robust inference that can be obtained by such a method is that GCs are on average $\alpha$ enhanced.
Namely, that $\alpha$-enhanced SSPs reproduce on average
the indices of Milky Way GCs better than solar-scaled SSPs.
It is important to note that the vast majority
of the globular clusters have measured indices in regions of the index-index diagrams poorly
explored by the several models (Mendel et al., 2007, c.f. their Fig. 2), therefore the inferred abundances are often
extrapolated.

Our goal is to calibrate observed Lick indices in Galactic globular clusters
for which detailed chemical abundances in stars have been recently 
measured with high resolution
spectroscopy. { In particular, we will make use of the average - over several works and several
stars in each globular cluster - abundances and abundance
ratios by Pritzl. et al. (2005) }
The aim of the project is to provide an empirical calibration relation in order to convert Lick
indices into abundances for a large range \emph{both in [Fe/H] (from -2.34 to
-0.06 dex) and [$\alpha$/Fe] (e.g. from -0.15 to 0.58 dex)}. 
The correlation between Lick indices such as $<Fe>$  (or Mgb) and [Fe/H]
has already been shown and discussed by other works, the latest and
most accurate being by Puzia et al. (2002), on a very similar range in [Fe/H].
With respect to these works, our calibration will be more robust in that
we will double the number of GCs - all of them observed by the same instrument
and with the same settings - and we do not limit the analysis to GCs
belonging to the Milky Way bulge (as in Puzia et al., 2002).
Furthermore, we use the latest [Fe/H] ratios derived from high resolution spectroscopy,
whereas previous works used a generic "metallicity", often based on older (photometric) measurements.

Finally, a careful study of the relation between indices and the abundance
of the main absorbing species at the relevant wavelengths in both our
observations and in synthetic spectra, will shed some light on the reason
of the results by Mendel et al. (2007) discussed above.

The plan of the paper is the following. We will discuss the observations and data reduction process in Secs.~\ref{obs} and~\ref{red}, respectively.
Our results will be presented in Sec.~\ref{res} and some implications are discussed in Sec. 5. Conclusions are drawn in Sec. 6.

\section{Observations}
\label{obs}
\begin{table}
\begin{flushleft}
\caption[]{List of Targets}
\begin{tabular}{l|llllllllllll}
\hline
Run & Globular cluster & $\alpha$&$\delta$ \\
\hline

B&NGC 104   &00 24 05.2& -72 04 57.9       \\
B&NGC 362   &01 03 14.3& -70 50 53.6       \\
A&NGC 3201  &10 17 36.8& -46 24 40.4       \\
A&M 68      &12 39 28.0& -26 44 34.9       \\
A&NGC 4833  &12 59 35.0& -70 52 28.6       \\
A&M 5       &15 18 33.7& +02 04 57.7       \\
A&M 80      &16 17 02.5& -22 58 30.4       \\
A&M 4       &16 23 35.4& -26 31 31.9       \\
A&M 12      &16 47 14.5& -01 56 52.1       \\
A&M 10      &16 57 09.0& -04 05 57.6       \\
A&NGC 6287  &17 05 09.3& -22 42 28.8       \\
A&NGC 6293  &17 10 10.4& -26 34 54.2       \\
A&NGC 6342  &17 21 10.1& -19 35 14.7       \\
A&NGC 6352  &17 25 29.1& -48 25 21.7       \\
A&NGC 6362  &17 31 54.9& -67 02 52.3       \\
A&NGC 6397  &17 40 41.4& -53 40 25.3       \\
A&NGC 6528  &18 04 49.6& -30 03 20.8       \\
A&NGC 6541  &18 08 02.2& -43 42 19.7       \\
A&NGC 6553  &18 09 15.7& -25 54 27.9       \\
A&M 22      &18 36 24.2& -23 54 12.2       \\
A&M 54      &18 55 03.3& -30 28 42.6       \\
A&NGC 6752  &19 10 51.8& -59 58 54.7       \\
A&M 30      &21 40 22.0& -23 10 44.6       \\
\hline
\label{list}
\end{tabular}
\end{flushleft}
\end{table}

We observed 23 Galactic globular clusters (see Table~\ref{list})
during two observing runs\footnote{run A:3 nights is May 2006 - run B:one night in September 2006}
at the ESO New Technology Telescope in La Silla using EMMI (Dekker et al. 1986). In particular,
we made use of the low-resolution RILD Grism 5, which covers a
wavelength range $\sim 380-700$ nm with a dispersion of 55 nm/mm,
with 1x1 binning and slow read-out to reduce the noise. One pixel corresponds to 0.166''.
The slit width was set to 2''. 
The resulting resolution is $\sim 10\AA$ FWHM, very close to the actual 
nominal $\sim 8.4-10 \AA$ resolution that
the Lick system has in the wavelength range relevant for our study. This set-up
has been chosen to minimise the correction otherwise needed to set the observed spectra
into the Lick resolution.

Spectra were taken with the slit in two different angles (East-West and North-South direction) for each cluster, to avoid
abnormal contribution from individual bright stars and to achieve
a secure sample of the underlying stellar population.

Individual exposures were adjusted to avoid saturation.
We separate our sample into two runs in order to ensure the observability
of each object at low air-masses. 
We estimated the exposures time by means of the 
ETC - Optical Spectroscopy Mode Version 3.0.6 - tool. In particular,
we used the central surface brightness in the V band provided by the 2003
update of the Harris (1996) catalogue for our sample of globular clusters and
required a S/N=50, for an airmass $\le 1.3$ and seeing below 2''.
To give an example, we found an exposure time of $\sim 100$ s for bright objects 
(i.e. V surface brightness $\sim 14-15\,\rm mag/arcsec^2$)
and $\sim 6800$ s for the faintest ones (i.e. V surface brightness $> 19\,\rm mag/arcsec^2$).
However, in order to reduce cosmic ray events, several shorter exposures
of each target were taken and then added together.
Lick standard stars 
were observed each night. The stars were slightly defocussed
in order to avoid saturation.

We observed under cloudy conditions during the first half of the first
night and for most of the last night. The science frames acquired
during these period have not been used in the following discussion.
Moreover, during the last night we had to discard some GCs in the original proposal list, since
they were no longer visible when the weather conditions improved.

\begin{table*}
\centering
\begin{minipage}{120mm}
\begin{flushleft}
\caption[]{Final corrected Lick indices - I}
\begin{tabular}{l|llllllllllll}
\hline
\hline
GC  &  H$_{\beta}  $&$\pm$&$  Fe5015  $&$\pm$&$  Mg_2  $&$\pm$&  Mg$_b$  &$\pm$&  Fe5270  &$\pm$\\
NGC104  &      1.66&     0.05&     3.65&     0.16&     0.164&     0.017&     2.56&     0.17&     2.07&     0.05\\
NGC362  &      1.87&     0.03&     2.60&     0.01&     0.089&     0.002&     1.00&     0.04&     1.41&     0.00\\
NGC3201 &      2.46&     0.39&     1.70&     0.41&     0.064&     0.002&     0.99&     0.03&     1.01&     0.07\\
M68     &      2.43&     0.16&     0.72&     0.09&     0.046&     0.000&     0.68&     0.01&     0.63&     0.08\\
NGC4833 &      2.31&     0.18&     1.26&     0.14&     0.055&     0.002&     0.50&     0.03&     0.80&     0.10\\
M5      &      2.65&     0.14&     2.05&     0.10&     0.079&     0.005&     1.11&     0.02&     1.29&     0.08\\
M80     &      2.31&     0.06&     1.52&     0.04&     0.059&     0.002&     0.89&     0.01&     0.99&     0.01\\
M4      &      2.37&     0.48&     2.93&     0.05&     0.119&     0.009&     1.71&     0.32&     1.63&     0.00\\
M12     &      2.58&     0.34&     1.55&     0.20&     0.073&     0.008&     1.21&     0.14&     1.03&     0.11\\
M10     &      2.74&     0.22&     1.48&     0.02&     0.064&     0.000&     0.86&     0.03&     0.93&     0.01\\
NGC6287 &      2.87&     0.27&     0.15&     0.19&     0.048&     0.000&     0.48&     0.07&     0.40&     0.05\\
NGC6293 &      2.80&     0.32&     0.54&     0.14&     0.048&     0.002&     0.62&     0.05&     0.60&     0.06\\
NGC6342 &      1.60&     0.06&     3.51&     0.22&     0.153&     0.001&     2.35&     0.08&     1.90&     0.11\\
NGC6352 &      1.42&     0.35&     4.12&     0.66&     0.198&     0.064&     2.91&     0.56&     2.36&     0.42\\
NGC6362 &      2.81&     0.63&     2.20&     0.33&     0.078&     0.006&     1.45&     0.08&     1.17&     0.09\\
NGC6397 &      3.03&     0.48&     0.89&     0.24&     0.048&     0.003&     0.63&     0.09&     0.77&     0.19\\
NGC6528 &      1.51&     0.13&     5.63&     0.00&     0.251&     0.005&     3.34&     0.05&     2.78&     0.16\\
NGC6541 &      2.65&     0.19&     1.04&     0.05&     0.056&     0.001&     0.74&     0.02&     0.93&     0.05\\
NGC6553 &      1.84&     0.20&     5.63&     0.22&     0.242&     0.014&     3.66&     0.04&     2.68&     0.04\\
M22     &      2.63&     0.07&     1.38&     0.45&     0.071&     0.020&     1.03&     0.13&     0.92&     0.15\\
M54     &      2.37&     0.02&     2.68&     0.00&     0.086&     0.000&     1.07&     0.01&     1.65&     0.01\\
NGC6752 &      2.53&     0.26&     1.50&     0.40&     0.064&     0.003&     1.00&     0.01&     1.06&     0.11\\
M30     &      2.49&     0.14&     0.92&     0.42&     0.056&     0.015&     0.67&     0.09&     0.77&     0.11\\
\hline
\label{table3}
\end{tabular}
\end{flushleft}
\end{minipage}
\end{table*}

\begin{table*}
\centering
\begin{minipage}{120mm}
\begin{flushleft}
\caption[]{Final corrected Lick indices - II}
\begin{tabular}{l|llllllllllll}
\hline
\hline

GC  &$  Ca4227  $&$\pm$&$  Ca4455  $&$\pm$&$  NaD  $&$\pm$&  TiO$_1$  &$\pm$&  TiO$_2$  &$\pm$\\
NGC104  &      0.74&     0.094&     0.79&     0.05&     1.80&     0.033&     0.0192&     0.0021&     0.0352&     0.0073\\
NGC363  &      0.68&     0.086&     0.84&     0.22&     0.73&     0.005&     0.0103&     0.0033&     0.0170&     0.0099\\
NGC3201 &      0.37&     0.085&     0.22&     0.02&     1.84&     0.135&     0.0073&     0.0005&     0.0104&     0.0036\\
M68     &      0.09&     0.004&    -0.04&     0.04&     1.12&     0.047&     0.0060&     0.0008&     0.0035&     0.0022\\
NGC4833 &      0.21&     0.017&     0.09&     0.05&     1.72&     0.069&     0.0054&     0.0004&     0.0069&     0.0018\\
M5      &      0.32&     0.007&     0.31&     0.03&     1.09&     0.018&     0.0129&     0.0001&     0.0032&     0.0011\\
M80     &      0.30&     0.043&     0.34&     0.09&     1.47&     0.034&     0.0140&     0.0007&     0.0031&     0.0000\\
M4      &      0.29&     0.108&     0.16&     0.15&     1.49&     0.195&     0.0076&     0.0001&     0.0099&     0.0017\\
M12     &      0.23&     0.053&     0.15&     0.01&     1.51&     0.014&     0.0143&     0.0006&     0.0064&     0.0001\\
M10     &      0.23&     0.010&     0.10&     0.03&     1.19&     0.018&     0.0043&     0.0010&     0.0032&     0.0003\\
NGC6287 &      0.03&     0.032&    -0.22&     0.06&     2.13&     0.090&     0.0043&     0.0022&     0.0056&     0.0019\\
NGC6293 &      0.06&     0.049&    -0.01&     0.04&     2.17&     0.027&     0.0019&     0.0003&     0.0025&     0.0011\\
NGC6342 &      0.75&     0.048&     0.86&     0.08&     2.78&     0.018&     0.0074&     0.0018&     0.0133&     0.0009\\
NGC6352 &      0.36&     0.007&     0.27&     0.49&     3.68&     0.008&     0.0084&     0.0010&     0.0169&     0.0050\\
NGC6362 &      0.43&     0.183&     0.07&     0.03&     1.78&     0.431&     0.0129&     0.0061&     0.0087&     0.0042\\
NGC6397 &      0.10&     0.039&     0.22&     0.20&     1.47&     0.013&     0.0081&     0.0006&    -0.0004&     0.0001\\
NGC6528 &      0.97&     0.090&     1.79&     0.03&     5.01&     0.099&     0.0321&     0.0012&     0.0673&     0.0033\\
NGC6541 &      0.11&     0.013&    -0.01&     0.05&     1.95&     0.036&     0.0045&     0.0008&     0.0003&     0.0004\\
NGC6553 &      0.90&     0.039&     1.07&     0.15&     3.53&     0.042&     0.0218&     0.0025&     0.0379&     0.0027\\
M22     &      0.09&     0.003&     0.04&     0.02&     2.23&     0.130&     0.0040&     0.0017&     0.0072&     0.0044\\
M54     &      0.45&     0.046&     0.63&     0.00&     1.48&     0.016&     0.0069&     0.0005&     0.0155&     0.0004\\
NGC6752 &      0.21&     0.006&     0.15&     0.05&     0.82&     0.053&    -0.0002&     0.0002&    -0.0003&     0.0007\\
M30     &     -0.00&     0.002&     0.10&     0.00&     0.88&     0.031&     0.0047&     0.0051&    -0.0082&     0.0067\\

\hline
\label{table3bis}
\end{tabular}
\end{flushleft}
\end{minipage}
\end{table*}

\begin{table*}
\centering
\begin{minipage}{120mm}
\begin{flushleft}
\caption[]{H$_{\alpha}$ and Serven et al.'s indices}
\begin{tabular}{l|llllllllllll}
\hline
\hline

GC  &$  Ba4552  $&$\pm$&$  Eu4592  $&$\pm$&$  H_{\alpha}  $&$\pm$\\

NGC104  &      0.7408&     0.0945&     0.7951&     0.0573&     1.8041&     0.0334\\
NGC363  &      0.6823&     0.0869&     0.8431&     0.2239&     0.7398&     0.0058\\
NGC3201 &      0.3730&     0.0852&     0.2294&     0.0294&     1.8414&     0.1352\\
M68     &      0.0934&     0.0045&    -0.0479&     0.0409&     1.1214&     0.0472\\
NGC4833 &      0.2177&     0.0175&     0.0988&     0.0568&     1.7210&     0.0691\\
M5      &      0.3234&     0.0075&     0.3190&     0.0360&     1.0958&     0.0180\\
M80     &      0.3010&     0.0438&     0.3412&     0.0947&     1.4735&     0.0347\\
M4      &      0.2957&     0.1082&     0.1661&     0.1519&     1.4967&     0.1954\\
M12     &      0.2390&     0.0535&     0.1557&     0.0131&     1.5193&     0.0147\\
M10     &      0.2376&     0.0108&     0.1000&     0.0397&     1.1910&     0.0187\\
NGC6287 &      0.0324&     0.0320&    -0.2268&     0.0684&     2.1397&     0.0908\\
NGC6293 &      0.0692&     0.0494&    -0.0075&     0.0442&     2.1769&     0.0275\\
NGC6342 &      0.7527&     0.0488&     0.8659&     0.0801&     2.7854&     0.0189\\
NGC6352 &      0.3694&     0.0072&     0.2741&     0.4902&     3.6868&     0.0085\\
NGC6362 &      0.4304&     0.1835&     0.0778&     0.0324&     1.7833&     0.4317\\
NGC6397 &      0.1043&     0.0392&     0.2273&     0.2058&     1.4738&     0.0132\\
NGC6528 &      0.9788&     0.0902&     1.7906&     0.0374&     5.0128&     0.0996\\
NGC6541 &      0.1198&     0.0137&    -0.0074&     0.0545&     1.9591&     0.0369\\
NGC6553 &      0.9086&     0.0398&     1.0769&     0.1589&     3.5347&     0.0420\\
M22     &      0.0947&     0.0034&     0.0416&     0.0248&     2.2392&     0.1309\\
M54     &      0.4507&     0.0468&     0.6317&     0.0091&     1.4880&     0.0168\\
NGC6752 &      0.2156&     0.0061&     0.1519&     0.0550&     0.8221&     0.0531\\
M30     &     -0.0054&     0.0023&     0.1008&     0.0042&     0.8856&     0.0319\\

\hline
\label{table3ter}
\end{tabular}
\end{flushleft}
\end{minipage}
\end{table*}

\section{Data reduction}
\label{red}

We performed the standard data reduction steps by means of the ESO MIDAS software.
In particular, dark current and bias were removed and the images
were \emph{flat-fielded} by means of calibration frames acquired every night.
Science frames were cleaned from cosmic rays and bad pixels, then calibrated
in wavelength by means of a He-Ar lamp frame and finally re-binned. Wavelength calibration
was checked on the sky lines of science frames. Sky lines
were also used to estimate the (negligible) variable vignetting along the slit.
The sky spectrum was estimated from regions at the edges of the science
frames. The variability of the sky lines and stellar crowding, in fact, hampered us from using
separate sky frames that we took during the observing runs in a homogeneous
way for all clusters.
A comparison between the two methods is presented in the Appendix (Table~\ref{table5}).
The actual resolution of the observations ($\sim 10 \AA$) was
confirmed from the width of both sky lines and calibration frames.
Finally, after correction for the atmospheric extinction, one-dimensional integrated spectra were created by carefully 
avoiding bright stars and by taking only the central $\sim$ 1.5 arcmin (along the slit) of the entire
frame, in order to maximize the S/N ratio.
The typical value for the S/N per pixel at the central wavelength is $\sim$50.
At each step of the reduction process we also propagated the statistical error starting from the Poisson
noise of the science frame, the read-out-noise of EMMI and the 
sky subtraction as in Carollo et al. (1993, their Eq. 1).
We thus created an error spectrum. 

Line-strength indices were measured by means of a suitable routine ($lick\_ew$)
in the $EZ\_Ages$ package 
(Graves \& Schiavon, 2008). The code measures all 
Lick indices - according to the Trager et al. (1998) definition.
{ We modified the routine to also measure indices as defined
by Serven et al. (2005) and the H$_{\alpha}$ index as defined
in Cohen et al. (1998)}. 
The error spectra are used to calculate the error associated to each measured
index as in Cardiel et al. (1998, c.f. their Eq. 20). 
We repeat such a procedure for the two slit positions for each globular cluster.
The reader interested
in these intermediate steps in referred to the Appendix, where
we show examples for the most widely used indices 
in Table~\ref{table1} along with their statistical errors.
As expected (e.g. Cardiel et al., 1998), the high S/N
for our spectra implies statistical uncertainties of the
order of a few percent in the majority of the cases.
For each cluster we show the measurements in both EW and NS
directions along with their statistical uncertainties.
Whilst for the majority of the observed objects
the values for a given index taken along the two directions of the same cluster are very close,
they might disagree
at more than 3$\sigma$ level if only statistical errors were taken into
account. Such variations mirror the different sampling of the stellar light
in different positions of the same cluster and give an idea of the \emph{intrinsic} spread within one cluster.
We took the average of the two measurements as the representative
value of that index for a given cluster. We will make use of the difference between
the two directions as the estimate of the \emph{uncertainty}
associated to a given index. In particular, we use half
of the difference as an estimate of the error.
Such values are shown in Table~\ref{table2}.
We applied small corrections if the spectra are at resolution 
(as in our case) lower than the Lick/IDS system.
Since our resolution is about the same as the Lick system such
corrections are minimal (c.f. examples in Table~\ref{table6bis}).

As far as the Lick indices are concerned, the final step requires us to set our measurements on the Lick system.
In practice, reduction steps such as the wavelength calibration, the smoothing
of the spectra, leave always some residual offset from the standard reference
frame set by the Lick group (see Worthey et al., 1992).
The typical way to tackle the issue is to observe and reduce
Lick standard stars (Worthey et al., 1994)\footnote{Available at: http://astro.wsu.edu/worthey/html/system.html} and compare
the measured indices with those published by the Lick group (e.g. Worthey
et al., 1992). If some systematic offset is present it is
common to correct the observed indices in order
to \emph{set them} on the Lick system.
For the more relevant indices, we found that the measured
$H_{\beta}$ values for the Lick standard stars that we observed
are consistent, within the errors, with the values
given by the Lick group. Therefore no correction
has been applied. The same holds for the indices Fe5015, { Ca4227, TiO$_1$ , TiO$_2$, and NaD}.
Small corrections were instead applied to the indices
{ $Mg_2$ (0.024 mag), $Mg_b$ (0.12 \AA) and Fe5270 (0.17 \AA). 
 We found that a correction that depends on the index
strength was necessary for the index Fe5335. 
We therefore discarded it from the discussion.}
A comparison with the literature (e.g. Puzia et al., 2002)
demonstrates that our adopted corrections are very similar
to those in previous works.
{ While Cohen et al. (1998) measured their indices
at a resolution of $\sim$8\AA, the Serven et al. (2005) indices
have been introduced and tested in relation to a massive elliptical
galaxy with 200 km/s velocity dispersion. We chose to 
present the measurements for these indices in our native resolution, without any
further correction. }
{ The sample of final \emph{corrected} most widely used Lick indices is presented in Tables~\ref{table3}
and~\ref{table3bis}, whereas Table~\ref{table3ter} shows some non-Lick indices.
Other Lick and Serven et al.'s indices are available
upon request.}
We do not associate an error to the above mentioned
procedure, and we give the final error as the difference
between the two directions.
We then compared our measurements with available
indices from Puzia et al. (2002 and references therein), Trager et al. (1998) and Graves \& Schiavon (2008).
the typical differences between our results and the literature being $\sim$10\%.
When the same GC has been observed by more than one author,
it is remarkable that the differences between authors and between one author
and us are comparable to the differences between the measurements along the EW and NS directions
taken by us.

\section{Results}
\label{res}

In this section we present the main result of this project, namely
an empirical calibration between observed indices and abundances (and
abundance ratios) measured in the stars. After a brief review of the
index-index properties of our GC sample, we will start with
the most widely used metallicity indicator, i.e. [Fe/H]. Then,
we will study the Mg, Ca, Ti, Eu and Ba abundances.

\subsection{Index-index diagrams}

\begin{figure}
\includegraphics[width=8.5cm,height=9cm]{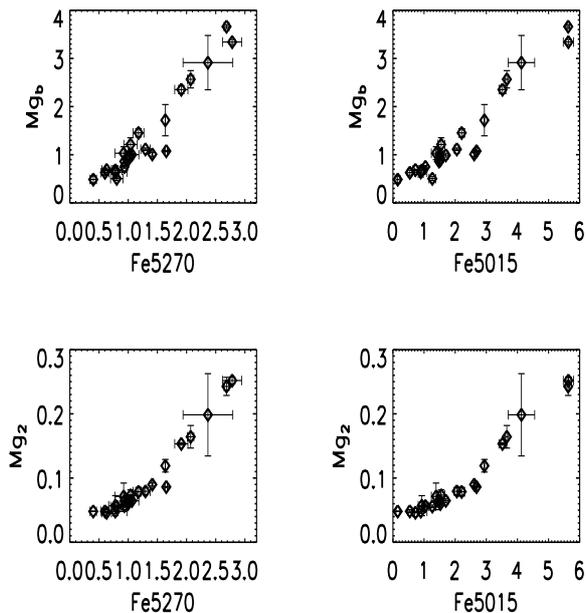}
\caption{Relation between observed Mg- and Fe-related indices. }
\label{fig3}
\end{figure}

In Fig.~\ref{fig3} we present the relation between observed Mg- and Fe-related indices
measured in this work. 

The reader should note that 
 the  $Mg_b$-Fe5270 relation tends
to deviate more and more from the 1:1 relation as the Fe5250 (and hence the metallicity) increases,
\emph{does not mean} that the $\alpha$-enhancement increases as well, as we will
see in the remainder of the paper.
In fact, as expected from chemical evolution studies of the Milky Way (e.g., Matteucci, 2001),
we know that [$\alpha$/Fe] decreases with [Fe/H], after a plateau, at [Fe/H]$\sim$-1. Such a trend is
evident also in entire sample of GCs by Pritzl et al. (2005).

\begin{figure}
\includegraphics[width=\linewidth]{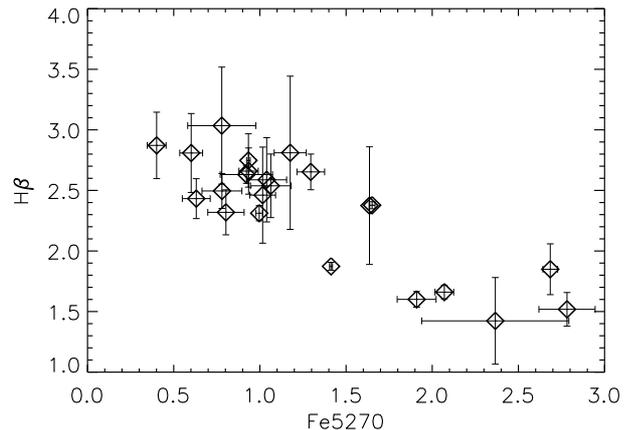}
\caption{Relation between observed H$_\beta$ and Fe5270 indices.}
\label{fig15}
\end{figure}

\begin{figure}
\includegraphics[width=\linewidth]{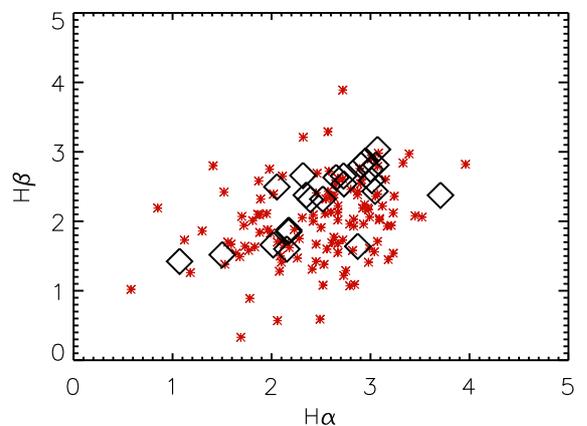}
\caption{Relation between observed H$_\beta$ and H$_\alpha$  indices. Diamonds: 
Milky Way globular clusters (this work); asterisks: M87 globular
clusters from Cohen et al. (1998).}
\label{fig15bis}
\end{figure}

\begin{figure}
\includegraphics[width=\linewidth]{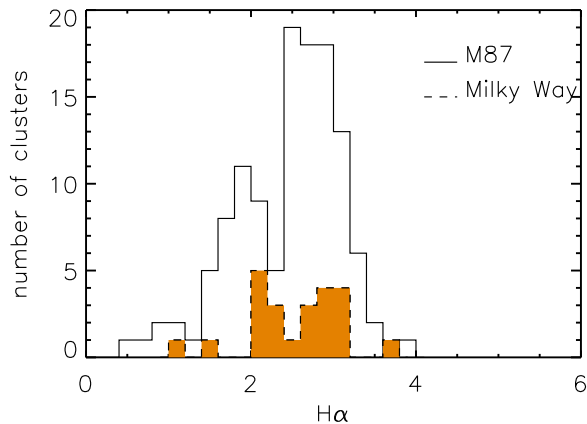}
\caption{Distribution for the values of the H$_\alpha$ index in M87 (solid, Cohen
et al., 1998) and the Milky Way (this work, dashed and shaded histogram).}
\label{fig15ter}
\end{figure}

In Fig.~\ref{fig15}, we show that higher values for the H$_\beta$ index correspond to lower values for metallicity-related indices.
This is not unexpected (see also Puzia et al. 2002, Burstein et al. 1984): the more metal poor, the
bluer the horizontal branch (e.g. Schiavon et al., 2004, Lee et al., 2009).
Here we do not revisit the issue and only conclude
that this hampers the use of the H$_\beta$ index as a pure age indicator, 
owing to its dependence on the horizontal branch morphology.
For the sake of the following discussion, it is important to note 
that models (e.g. Lee et al., 2009) show that the horizontal branch has little (if any)
effect on the metal indices that we study.
{ On the other hand, H$_\beta$ exhibits an almost 1:1 correlation with H$_\alpha$ (squares
in Fig.~\ref{fig15bis}).
Such a relation is tighter and somewhat steeper than the one found by using the M87 globular cluster
by Cohen et al. (1998) (asterisks in Fig.~\ref{fig15bis}).
However, the distribution in the values of the H$_\alpha$ index in M87 and
our sub-sample of Milky Way globular clusters are remarkably similar (Fig.~\ref{fig15ter}).
We refrain from a further interpretation given the large difference in size between
the two samples. 
We note that the H$_\alpha$ index has been shown by Serven et al. (2010) to provide
a useful independent estimate to correct the H$_\beta$ index for emission in galaxies.
Similarly, Poole et al. (2010) estimate the contamination from active
M dwarfs to their Milky Way globular cluster spectra.}

\begin{table}
\begin{flushleft}
\caption[]{$[X/H]=a\,\cdot Index +b$ relations.}
\begin{tabular}{l|llllllllllll}
\hline
           &    a     &       b    &      r     \\
\hline
[Fe/H]     &          &            &            \\
Fe5270     &      0.78   &  -2.37  &   0.93     \\
Fe5015    &      0.38    & -2.24   &  0.94        \\

\hline
[Mg/H]&&&\\
$Mg_2$   &       8.98  &   -1.76  &   0.88 \\      
$Mg_b$   &     0.56  &   -1.87   &  0.90  \\    

\hline
[Ca/H]&&&\\
Ca4227  &    1.80156  &   -1.75641 &    0.871257 \\

\hline
[Ba/H]&&&\\
Ba4552  &    6.96474  &  -2.39027 &    0.710823 \\

\hline
[Eu/H]&&&\\
Eu4592  &    4.54426  &   -1.23959 &    0.649848 \\

\hline
\label{table4}
\end{tabular}
\end{flushleft}
\end{table}

\subsection{An empirical calibration for the GC [Fe/H]}

\begin{figure}
\includegraphics[width=\linewidth,height=10cm]{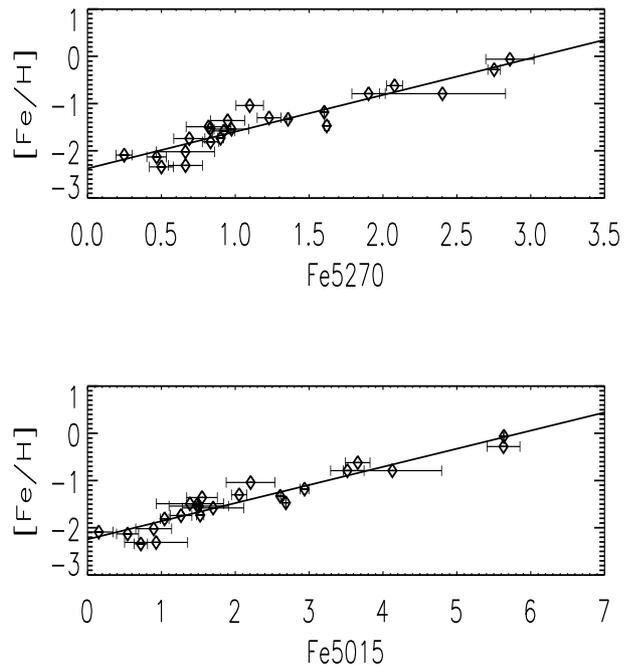}
\caption{Average [Fe/H] in stars by Pritzl et al. versus Fe-related Lick indices (this work).
The solid line is the formal linear regression.}
\label{fig1}
\end{figure}

In Fig.~\ref{fig1} we present the Fe-related index Fe5015 and Fe5270 versus
the { mean} Fe abundance in the stars of the GCs given by Pritzl et al. (2005).
The solid lines in Fig.~\ref{fig1} are obtained through a formal linear regression to the points
and the coefficients of the relations $[Fe/H]=a\,\cdot Index +b$ are given
in Table~\ref{table4} along with the correlation coefficients $r$.
{ We note that, in calculating these relations we implicitly kept the age
fixed. Therefore we warn the reader not to blindly use these calibrations
in external galaxies, where they can lead to an underestimation of the metallicity if
a younger (sub-)population of GCs exists.}
As expected, very tight linear relations link the Fe-indices to the
[Fe/H] abundance. Note that, even if several GCs show hints for multiple stellar populations,
the Fe content of their stars is highly homogeneous to within $\sim$ 10\% (Carretta et al., 2009).
A very similar calibration can be obtained if one uses the metallicity scale by Zinn \& West (1984) as
in the Harris (1996) catalogue. In particular, we find $[Fe/H]= 0.76 \cdot Fe5270 - 2.22$.
Recently, a new metallicity scale based on high resolution spectra (Carretta et al., 2009) has been
released. We have 9 GCs in common with Carretta et al.'s sample. Their new [Fe/H] abundances are within
a few percent of the values from Pritzl et al. that we adopted in this paper, so we do not
expect significant variations in the calibration even if such a more recent and homogeneous metallicity
scale were adopted.

Our results are also in agreement with Puzia et al.(2002)'s findings.
However, two significant improvements are present: i) we do
not limit the analysis to Milky Way Bulge GCs; ii) the 
empirical calibration relation between indices
and abundances makes use of abundances measured
in stars by means of high resolution spectra, whereas
previous works were based on some metallicity scale
derived mostly from photometric colours.

\begin{figure}
\includegraphics[width=\linewidth,height=6cm]{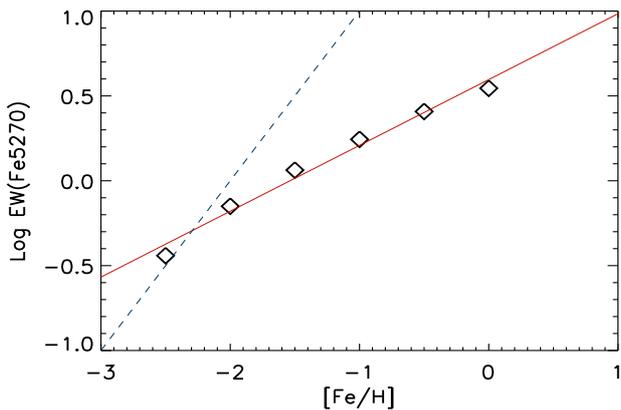}
\caption{Curve of growth-like diagram for the Fe5270 absorption as modelled
by the solar-scaled spectra of a K0 giant star.
The solid line is the formal linear regression. The dashed line gives the 1:1
relation as if the linear regime held up to [Fe/H]$\simeq -$2. }
\label{fig1a}
\end{figure}

Such a remarkable linear behavior between the index and the logarithm of the main
absorber might be explained in terms
of curve of growth. A qualitative explanation can be made by using synthetic spectra of a K0 giant
(P. Bonifacio, priv. comm.).
Below we will show that more detailed models of a SSP yield consistent results.
Indices like Fe5270 are measured in $\AA$ of equivalent width (EW),
therefore a curve of growth-like
diagram can be easily made by plotting, e.g. log (Fe5270) Vs. [Fe/H].
Fig.~\ref{fig1a} shows the curve of growth-like diagram
for our synthetic spectra, the solid line being a formal linear fit and the dashed
one giving the 1:1 relation.
Overall, we find log (Fe5270) $\sim 0.4\times$[Fe/H], suggestive of a  situation
where we are abandoning the linear regime (where log EW$\sim 1\times$ Log Abundance, dashed line)
at [Fe/H]$\sim$-2 and we enter the logarithmic saturated one (where EW$\sim$ Log Abundance,
hence $Index \sim$[Fe/H]). Such a conclusion is corroborated by inspection
of the lines in the high resolution synthetic spectrum and also holds for the Fe5015 index.
Also, the reader should note that, according to Tripicco \& Bell (1995), this index seems to be sensitive
to Mg, Ti and the total metallicity Z rather than Fe.
An important caveat applies
to the discussion: strictly speaking the curve of growth 
as a function of abundance is for a single absorption line, whereas
the EW of each Lick index includes the contribution from several lines,
not all related to the most important absorbing species at those wavelength. Namely
Fe and Mg indices are sensitive to variations in, e.g., Ca, C, Ti abundances (e.g. Thomas
et al., 2003, Lee et al. 2009). { Moreover, since the true spectroscopic continuum is lost at low
resolution, and since the index definition pseudocontinua bands will
fail to recover it, there will always be some shift and slope in index
values compared to a true curve-of-growth analysis.}

We note that these different regimes of the curve of growth were not taken into account
in the calculation of $\alpha$-enhanced indices as in Thomas et al. (2003). Instead,
the linear regime was a common assumption. An assessment of the error is beyond
the scope of the paper, and likely unnecessary because in more recent models (e.g. Lee et al., 2009)
the variation in the index is derived from the analysis
of an extensive library of synthetic spectra \emph{and} isochrones made for several chemical compositions. 
The use of Mg-enhanced composition for generating the spectra do not alter such
conclusions.

\subsection{An empirical calibration for the GC [Mg/H]}

\begin{figure}
\includegraphics[width=\linewidth,height=10cm]{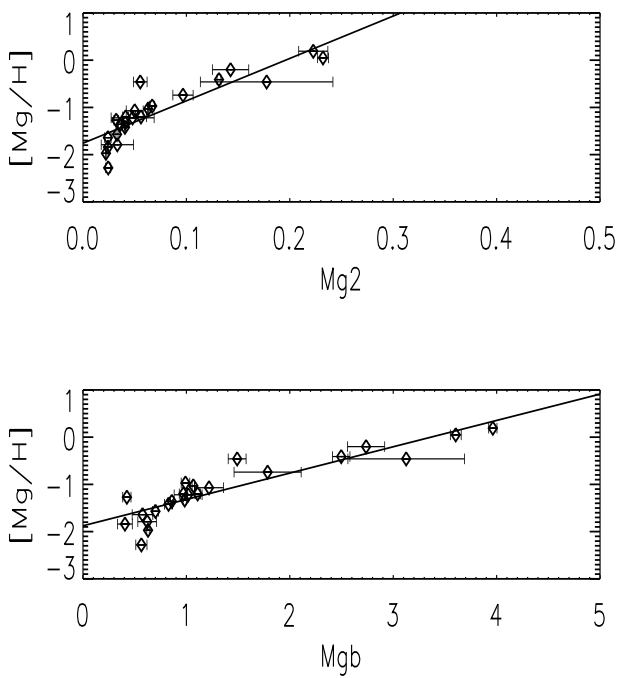}
\caption{Average [Mg/H] in stars by Pritzl et al. versus Mg-related Lick indices (this work).
The solid line is the formal linear regression.}
\label{fig4}
\end{figure}

A similar analysis with the Mg-related indices (Fig.~\ref{fig4}) shows
that [Mg/H] scales with $Mg_2$ and $Mg_b$. Also they roughly scale as the \emph{metallicity}. 
However at low values for the indices, the relations deviate from a straight line.

\begin{figure}
\includegraphics[width=\linewidth,height=6cm]{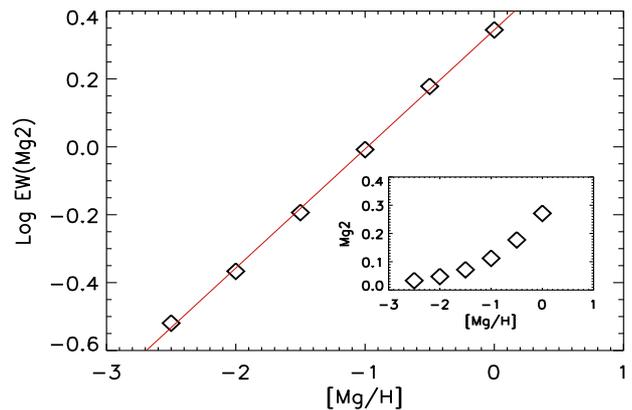}
\caption{Curve of growth-like diagram for the EW($Mg_2$) as modelled
by the solar-scaled spectra of a K0 giant star.
The solid line is the formal linear regression. The inset shows the Mg$_2$ index
as a function of the Mg abundance.}
\label{fig5a}
\end{figure}

We now try to explain the reason for this loss of sensitivity
to changes in the Mg abundance (in the inset of Fig.~\ref{fig5a} we show the 
$Mg_2$ index predicted by our K0 giant spectrum as a function of the [Mg/H] abundance in the star).
Indeed, a quadratic relation between Mg-indices
and [Mg/H] (see also Puzia et al., 2002) fits better than the simple linear relation and it
is expected from the index-index diagrams presented in Sec. 4.1.
Again, we can understand this behavior in terms of curve of growth.
We first notice that the $Mg_2$ index is the only one in the subset of
Lick indices studied here that is defined in magnitudes. Moreover,
it is \emph{not} defined as the EW of the absorption features in the
$Mg_2$ bandpass. Therefore some algebra is required to derive
the EW from the measure of the $Mg_2$. The behavior of the EW($Mg_2$)
as a function of [Mg/H] (main panel of Fig.~\ref{fig5a}) arrives again in the flat region of the curve
of growth and explains the relation between $Mg_2$ and [Mg/H] as a result
of the non linear relation between $Mg_2$ and EW($Mg_2$).
The main conclusion, however, is that below [Mg/H]= -1.5, the 
$Mg_2$ index is not a good measure of either the Mg abundance or the total
metallicity.

A similar behavior (and similar main conclusion) applies to the $Mg_b$ index.
In this case, however, an inspection of the high resolution spectra tell
us that it is the competition between the Mgb lines 
and other metal lines (mostly from Ca, Ti, Fe) in the flanking bands that makes the index insensitive to abundance changes at [Mg/H] below -1.5.
This is because, while the former lines have cores that saturate at quite low [Mg/H], the latter saturate at a slightly higher values for
[Mg/H].
In practice, changes in the depth of the central absorption features seem to be
compensated by the changes in the pseudo-continuum.
The results are unchanged if one uses synthetic spectra of a star
with an Mg-enhanced composition.
These prediction show a remarkable qualitative
agreement with those derived for integrated spectra of a SSP by Lee et al. (2008). 
Therefore we argue that our discussion based on the scrutiny of a
single star can be extended to the general case of a SSP, at least as a
partial explanation.
Other studies (Maraston et al., 2003) showed that at low metallicities 
the Mg indices of a SSP tend to be dominated by the lower main sequence,
making them prone to be affected by changes in the IMF due to the
dynamical evolution of the GCs.
This is probably why the indices calculated in a single K0 giant star
are weaker than both those measured in our GC sample and those
predicted by a theoretical SSP with a Salpeter (1955) IMF.

\begin{figure}
\includegraphics[width=\linewidth,height=6cm]{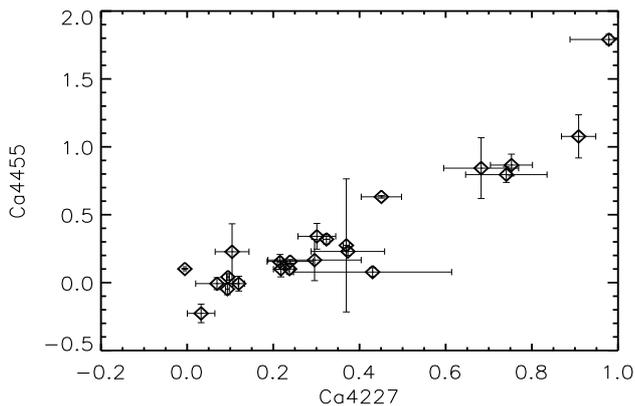}
\caption{Ca index-index diagrams.}
\label{fig8}
\end{figure}

\begin{figure}
\includegraphics[width=\linewidth,height=6cm]{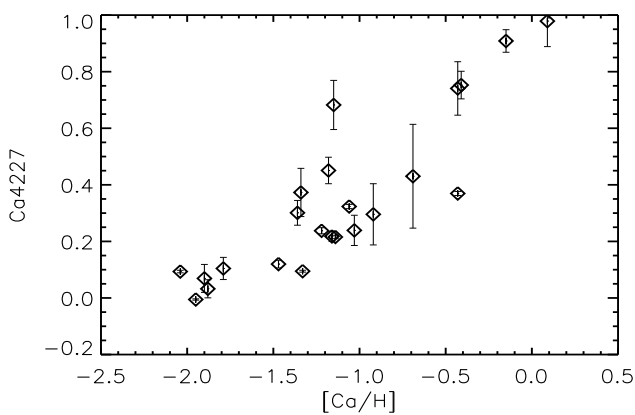}
\caption{Average [Ca/H] in stars by Pritzl et al. versus a Ca-related Lick index.}
\label{fig8bis}
\end{figure}

\subsection{An empirical calibration for other $\alpha$ elements}

Beside the well studied Mg and Fe indices, the high resolution data put us
in the position to provide - for the first time - empirical calibrations
for other elements. Let us start with the results for two
Ca-indices, namely Ca4227, Ca4455 (in the Lick system).
They track each other very well
(Fig.~\ref{fig8}) and correlate with the average Ca abundance measured in stars
as given by Pritzl et al. (2005), as shown in Fig.~\ref{fig8bis}.
The empirical calibrations are given in Table~\ref{table4}.
\begin{figure}
\includegraphics[width=\linewidth,height=6cm]{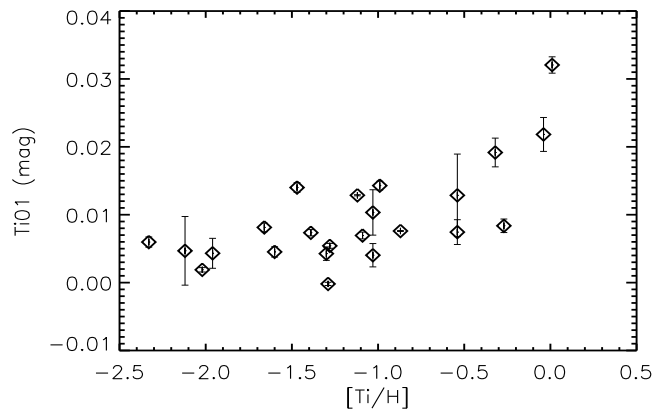}
\caption{Average [Ti/H] in stars by Pritzl et al. versus a Ti-related Lick index.}
\label{fig9}
\end{figure}
Ti is another element commonly enhanced as other $\alpha$ elements.
Here we note that the TiO$_1$ index trend with [Ti/H] is not linear, since
the index is measured in magnitudes (as we have seen for the $Mg_2$). 
Therefore we do not provide a linear fit for the empirical calibration.
We have found that the index TiO$_2$ closely tracks
TiO$_1$, therefore we have a similar trend with [Ti/H] (not
shown here).
As we will briefly discuss below for the Mg, none of these other enhanced elements offers
a simple way to infer a calibration for the [$\alpha$/Fe] ratio.

\subsection{An empirical calibration for neutron rich elements}

\begin{figure}
\includegraphics[width=\linewidth,height=6cm]{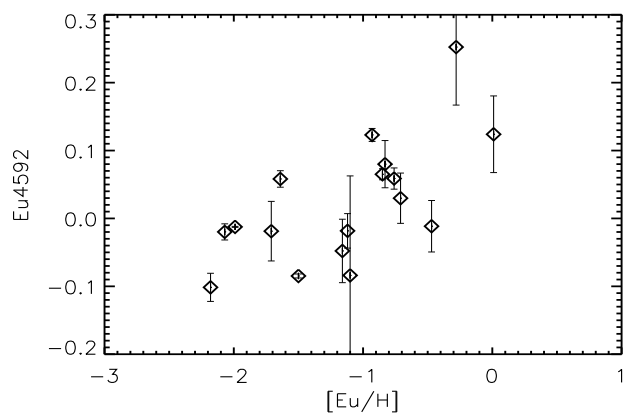}
\caption{Average [Eu/H] in stars by Pritzl et al. versus a Eu-related index from Serven et al. (2005).}
\label{fig10}
\end{figure}

\begin{figure}
\includegraphics[width=\linewidth,height=6cm]{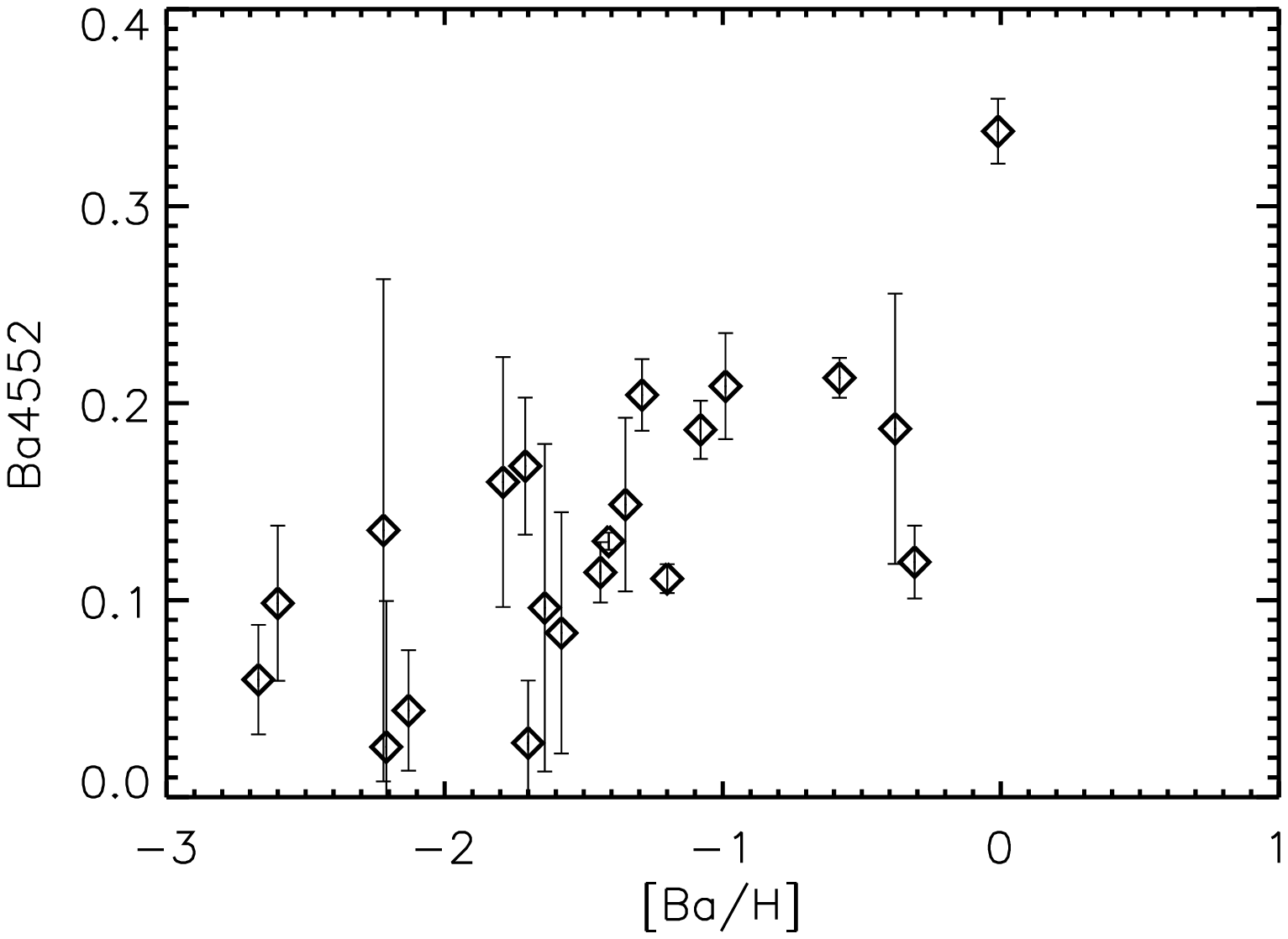}
\caption{Average [Ba/H] in stars by Pritzl et al. versus a Ba-related index from Serven et al. (2005).}
\label{fig11}
\end{figure}

Serven et al. (2005) also defined indices that allow the study of neutron rich elements.
In this section we adopt two of them to study the abundance of a typical
r-process element (Eu) and of a typical s-process element (Ba).
In Fig.~\ref{fig10} we show the results for the
Eu index, Eu4592, from Serven et al. (2005).
The results for Ba4552 are shown in Fig.~\ref{fig11}. The empirical calibrations
are given in Table~\ref{table4}.
We note that the relation involving Ba and Eu are less tight than the
ones for Fe and Mg.
This is due to the fact that the Serven et al. indices
are designed for very high ($>$100) S/N data, whereas our data
typically have S/N at most 40 at the relevant wavelengths. Nonetheless,
finding such a correlation is important because, to our knowledge,
this is the first time that these indices are tested on such a large metallicity range.
However, we stress that further work will be required to demonstrate that at our resolution
and S/N the contribution from other metals (mainly Fe) is negligible and that
the correlations shown in the figures is really due to an abundance increase.

\section{Discussion}

One further step in our analysis could have been to
further transform the indices into [Fe/H] abundances by means of the standard inversion
technique\footnote{That is a minimization technique that yields the best
set of abundances and age for a given set of measured indices and a specific SSP model}. Although we employed the $EZ\_Ages$ package, we cannot use it
because the metallicity grid (based on Schiavon, 2007, tracks) on which $EZ\_Ages$ in based
does not allow inversion at [Fe/H] below -1.3 and -0.8 for the solar-scaled
and the $\alpha$-enhanced cases, respectively, whereas, according to Pritzl et al.
most of our GCs are below these limits.

As for the use of other SSPs, such as the ones by Thomas et al. (2003, and their further improvements),
we refer to Mendel et al. (2009) and Lee et al. (2009) who did a thorough testing on GC data and
found a reasonable agreement between the SSP-inferred [Fe/H] and 
metallicity estimates based on resolved stellar populations (the former author) and
GC metallicity scales (the latter).
We note that almost all the tracks tested by Mendel et al. (2007)
to transform the line-strength indices into abundance ratios
fail to give a suitable $\alpha$-enhancements as a function of Fe abundance, namely they cannot 
reproduce the observed [Mg/Fe]-[Fe/H] relation
observed in the stars of the same GCs (Pritzl et al., 2005 - see Fig. 6 in Mendel et al. 2007).
Whether this is a problem intrinsic to the SSP libraries or it is rooted in the
inversion technique it has to be understood.
Such a problem it is somewhat expected  from the above analysis. We found that [Fe/H]$<-$1 is extremely difficult
to discriminate between a track pertaining to the solar composition and one built
assuming a +0.4 dex enhancement in the [Mg/Fe] ratio.

Our findings { eventually provide an explanation for the difficulties} by other works.
For instance, Puzia et al. (2005) found it difficult to discriminate among $\alpha$-enhanced and solar
scaled (extra-galactic) GCs at below $Mg_2\sim$0.2 mag and $<$Fe$>$ below 2$\AA$. By using the relations
derived in the previous sections (e.g. Fig.~\ref{fig1}), these values
correspond to [Fe/H]$<$-1, i.e. where the theoretical curves in the index-index plane
come closer and closer. { Similarly, this happens in the recent update of the Thomas
et al. (2003) stellar population models (Thomas et al., 2010).}

Clearly, if theoretical models with different $\alpha$-enhancement differ so little,
errors in the measurements may render impossible the derivation of the true [Mg/Fe] ratio
in the [Fe/H]$<$-1 regime.
This has interesting consequences, since high resolution spectra from GC stars show a typical level of
$\alpha$-enhancement (0.3 dex) comparable (and in a few cases even higher) than those typical
of massive ellipticals.

\begin{figure}
\includegraphics[width=\linewidth,height=6cm]{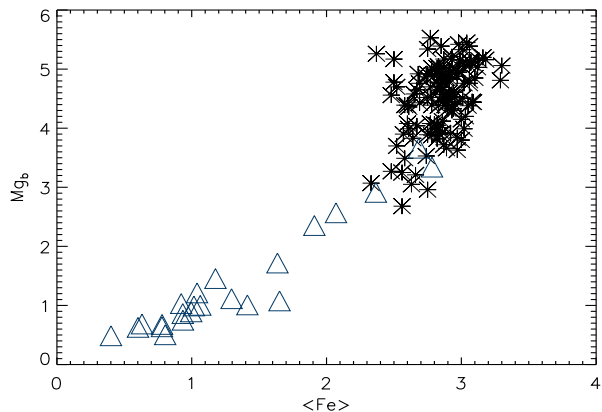}
\caption{Position in the index-index diagram of our GCs (triangles) and a sample
of elliptical galaxies (asterisks, Thomas et al., 2005).}
\label{fig77}
\end{figure}

For elliptical galaxies, although the relation between Mg-related indices and Fe-related ones
is a continuation of the overall trend seen in GC (Fig.~\ref{fig77}), it suddenly becomes
much steeper (Burstein et al.,1984; Worthey et al., 1992). The original interpretation (Burstein et al., 
1984) was that something else ([Mg/Fe] enhancement) is contributing to the galaxy
$Mg_b$ excess. 
Whilst GCs can be considered SSPs, elliptical galaxies cannot.
They are a Composite Stellar Population made of several SSPs.
This implies for the latter class that the location in the $Mg_b$-Fe5270 diagram (or equivalent)
is somewhat a measure of the average value among several SSPs that formed during
the galactic evolution.

\begin{figure}
\includegraphics[width=\linewidth]{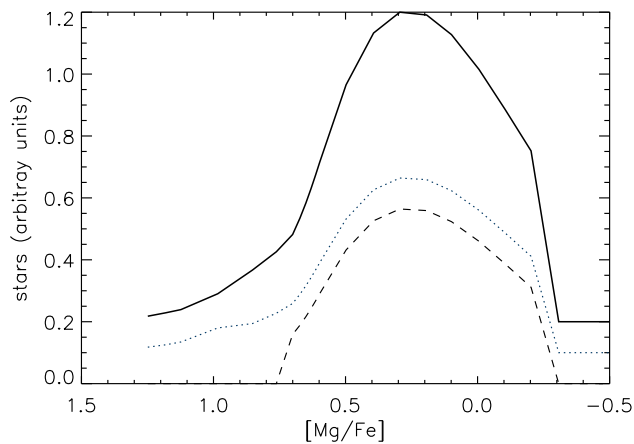}
\caption{Stellar mass distribution as a function of [Mg/Fe] as predicted by the PM04 chemical evolution model for a typical
elliptical galaxy (solid line). The luminosity distribution as a function of [Mg/Fe] is given by the dotted line. The dashed
line gives the luminosity distribution when the SSPs formed at [Fe/H]$<-$1 are not taken into account (see text). The curves
have been arbitrarily rescaled to the same value and offset in order to make differences in the tails more visible.}
\label{fig66}
\end{figure}

We expect the first SSPs to form in these galaxies
to contribute more to the total light due to the higher blue/optical luminosity with respect to later SSPs\footnote{
Line blanketing in metal richer populations suppress the flux in the wavelengths
where the spectra are taken. Also, note
that the average ages and the average $\alpha$ enhancement imply that
the SSPs are rather old (above 10 Gyr) and that the spread in ages
in a single galaxy should not exceed $\sim$ 1 Gyr. Therefore we
neglect age effects in this discussion.}.
In other words, a Composite Stellar Population  must have a higher average metalliticy of a SSP featuring the same line-strength
index (Greggio, 1997), because the lowest metallicity tail of the stellar metallicity
distribution has a non-negligible role in the integrated spectrum.
In the light of our results on the GCs we expect the first SSPs 
to have little impact on the Mg-related index. In the sense that, even
if they are the most $\alpha$-enhanced SSPs in the galaxy, since they
formed only out of SNII ejecta, their Mg-related indices will be
fairly low and indistinguishable from the values of a solar-scaled
SSP. Therefore, one may tempted to say that the $\alpha$-enhancement
that we ``measure'' in galaxies is lower than the true $\alpha$-enhancement.
With the help of Fig.~\ref{fig66} we show that this is (luckily) not the case.
The solid line is the stellar mass distribution as a function of [Mg/Fe] as predicted by the Pipino \& Matteucci (2004, PM04)
chemical evolution model for a typical massive elliptical galaxy. The mass-weighted average of such a distribution
is $<$[Mg/Fe]$_{true,*}>=$ 0.43 dex. This is the \emph{true} average that an observer would wish to obtain from the inversion
of the indices into abundances. In reality, the observed quantity is a luminosity weighted value.
We convert the stellar mass distribution into the stellar  
luminosity distribution as a function of [Mg/Fe] (dotted line) by using the M/L ratio for
a 12 Gyr old SSP with Salpeter IMF as a function of [Fe/H] from Maraston et al. (2003).
The predicted luminosity average is $<$[Mg/Fe]$_{true,lum}>=$ 0.45 dex.
We now assume that not all the SSPs that make our galaxy contribute to the measurement of the
$\alpha$-enhancement. In particular, the dashed
line gives the luminosity distribution when the SSPs formed with [Fe/H]$<-$1 are not taken into account.
We chose [Fe/H]= -1 as the limiting boundary because we showed that below such a limit
the [Mg/Fe] does not make any difference in the predicted Mg- related indices.
The average [Mg/Fe] inferred from this last distribution is $<$[Mg/Fe]$_{obs}>=$ 0.38 dex.
This exercise shows that the [Mg/Fe] will only be \emph{underestimated} by a modest amount (0.05 dex).
A lower mass elliptical, with a more extended and quieter star formation history than
the example displayed in Fig.~\ref{fig66} might have a larger proportion of SSPs formed with [Fe/H]$<-$1
and hence a larger underestimation of its $<$[Mg/Fe]$_{true}>$. A further analysis
would require a proper weighting of the fluxes of all the stars involved according
to their spectral type, which is beyond the scope of the present paper.

Finally, we note that Mg is one of the elements that show signs of large star to
star variation in (anti)correlation
with Al(e.g. Carretta et al., 2010 and
references therein, Gratton et al., 2004) due to self-enrichment from AGB stars (e.g. D'Antona \& Ventura, 2007) or rotating stars ( Decressin et al., 2007) and possibly related to the presence of at least two stellar generations
in most GCs (Carretta et al., 2009, D'ercole et al., 2009). 
Unfortunately, Al does not contribute to Lick indices, { and our S/N is not high
enough to have an accurate measurement of the Al3953 index defined by Serven et al. (2005),} therefore we have no
means of acquiring a deeper insight with the low-resolution integrated light spectroscopy. 
A promising tool may be high resolution integrated light spectroscopy (e.g. McWilliam \& Bernestein, 2008).

\section{Conclusions}

We carried out a long-slit spectroscopic study of 23
Galactic globular clusters for which detailed chemical abundances
in stars have been recently measured. We measured the metallicity indices as coded
by the \emph{Lick system}, Cohen et al. (1998) and by Serven et al. (2005) that we measured in low-resolution integrated spectra of the GC light.
We compared them to average abundances and abundance ratios in their stars taken from the compilation
by Pritzl et al. (2005) as well as to synthetic models.

We provided an empirical calibration relation in order to convert these
indices into abundances for a large range in the \emph{observed} [Fe/H] and [alpha/Fe]. 
The Mg- and Fe-metallicity calibrations supersede the previously adopted ones because they are be based
on the real abundance pattern of the stars instead of the commonly adopted
metallicity scale derived from, e.g., colours.
We present novel calibrations for Ca, Ti, Ba and Eu based on the same set of average
abundances in stars.

Below we summarize our main results:
\begin{itemize}
\item For the first time we present the H$_\alpha$ index (Cohen et al., 1998) for Milky Way globular clusters.
In our globular clusters H$_{\beta}$ tracks H$_{\alpha}$ very well. The distribution in values for H$_{\alpha}$
is similar to that inferred by Cohen et al. (1998) for M87.

\item Calibration of [Fe/H] as a function of Fe-indices is the best estimator of the GC metallicity
(also because the small star to star variation in [Fe/H]). The relation between indices and
[Fe/H] is linear to a good approximation at [Fe/H] $>-$2, when a regime similar to the
saturated regime of the curve of growth sets in.

\item Mg-indices are not reliable below [Mg/H]=-1.5 for effects due to the contribution
of metal lines in the pseudo continuum and, possibly, IMF effects.


\item At [Fe/H], [Mg/H] $<$ -1.5, it is impossible to measure the effect of [Mg/Fe] in the spectra.
At (slightly) higher metallicities, poor statistics, errors and star-to-star variation within
GCs hampered us from deriving a simple relation between [Mg/Fe] and (a combination of) indices.
Only in the very high metallicity regime of elliptical galaxies is it possible to safely
use Mg- and Fe-indices to infer the mean [Mg/Fe] in stars.

\item Since a small fraction of low-metallicity $\alpha$-enhanced stars do exist in elliptical 
galaxies, we estimate that SSP-equivalent value of the [Mg/Fe] might underestimate the
true average value in their stellar populations.

\item We show that the Lick Ca- and TiO- indices correlate with [Ca/H] and [Ti/H], respectively.
The relations are less tight than in the Mg case. Also these $\alpha$ elements do not allow the
construction of a reliable calibration for measuring the global $\alpha$-enhancement.

\item We show that the Serven et al. (2005) Eu and Ba indices correlate with [Eu/H] and [Ba/H], respectively,
although with a larger scatter than in the Mg- and Fe-metallicity calibration.
\end{itemize}

\section*{Acknowledgments} 
We warmly thank the referee, G. Worthey, for suggestions that improved the quality of the paper,
E.Pompei for her invaluable help during
the observing runs, S.Samurovic for having provided help with MIDAS at
the beginning of this work and P. Bonifacio for having provided synthetic spectra. 
We acknowledge useful discussions from G.Graves, M. Koleva, H.-C. Lee, 
M.Sarzi, A. McWilliam and P.Sanchez-Blazquez. 
This research has made use of: the SIMBAD database, operated at CDS, Strasbourg, France;
the Harris Catalog of Parameters for Milky Way Globular Clusters - Feb 2003 revision - available at 
http://physwww.mcmaster.ca/
the Lick/IDS data mantained by G. Worthey (http://astro.wsu.edu/worthey/html/system.html).

\clearpage

\begin{appendix}
\section{Sky subtraction}

In this Appendix we show two examples of the results for differing
methods of sky subtraction for a subset of the measured indices.
As discussed in the main text, night sky variability and the location
of some GCs in crowded fields did not allow us to always find
a suitable region close enough to the target that could be used
as an external sky frame.
When this was possible, the difference in the results with respect to the sky estimate
from the edges of the science frame is very small, especially in the case of metal lines.
We note here that for the same two clusters, Puzia et al. (2002) found a larger difference
between the two way of accounting for the sky background. This can be the consequence
of the target region chosen to measure the sky spectrum. However, based on our tests,
we agree with them that the \emph{internal} derivation of the background is more reliable.

\begin{table*}
\centering
\begin{minipage}{120mm}
\begin{flushleft}
\caption[]{Uncorrected Lick indices. Example of same frame sky estimate versus external frame}
\begin{tabular}{l|llllllllllll}
\hline
\hline
GC  &  H$_{\beta}  $&$\pm$&$  Fe5015  $&$\pm$&$  Mg_2  $&$\pm$&  Mg$_b$  &$\pm$&  Fe5270  &$\pm$\\
ext. frame \\
NGC6528EW  &      1.46&     0.12&     5.60&     0.24&     0.2514&     0.0029&     3.67&     0.11&     2.92&     0.11\\
NGC6528NS  &      1.32&     0.12&     5.74&     0.23&     0.2280&     0.0028&     3.59&     0.10&     2.75&     0.11\\
same frame \\
NGC6528EW  &      1.65&     0.07&     5.64&     0.14&     0.2375&     0.0018&     3.65&     0.06&     3.02&     0.07\\
NGC6528NS  &      1.38&     0.07&     5.62&     0.13&     0.2266&     0.0016&     3.55&     0.06&     2.69&     0.06\\
\hline
ext.frame \\
NGC6553EW  &      1.94&     0.06&     5.19&     0.12&     0.2130&     0.0015&     3.68&     0.05&     2.56&     0.06\\
NGC6553NS  &      1.41&     0.07&     4.77&     0.15&     0.1914&     0.0018&     3.70&     0.06&     2.50&     0.07\\
same frame \\
NGC6553EW  &      2.05&     0.04&     5.85&     0.07&     0.2368&     0.0009&     4.00&     0.03&     2.79&     0.03\\
NGC6553NS  &      1.63&     0.04&     5.40&     0.08&     0.2086&     0.0010&     3.92&     0.03&     2.70&     0.03\\
\label{table5}
\end{tabular}
\end{flushleft}
\end{minipage}
\end{table*}

\section{Uncorrected indices}

In this section we show the uncorrected Lick indices (after sky subtraction) measured
along the two directions for each individual cluster (Table~\ref{table1}) as
well as their average (Table~\ref{table2}). See the main body of the paper for details.
We recall that, for ``uncorrected'' we mean data that have not (yet) been corrected for any offset between our and the Lick
system due to, e.g., residuals in the sky subtraction, systematics in the wavelength calibration.
The example is limited to a subset of the measured indices, the remainder of the sample being available upon request.

\begin{table*}
\centering
\begin{minipage}{120mm}
\begin{flushleft}
\caption[]{Lick indices measured along EW and NS directions and their associated statistical uncertainties}
\begin{tabular}{l|llllllllllll}
\hline
\hline
GC  &  H$_{\beta}  $&$\pm$&$  Fe5015  $&$\pm$&$  Mg_2  $&$\pm$&$  Mg_b $&$\pm$&  Fe5270  &$\pm$\\
NGC104EW  &      1.60&     0.01&     3.82&     0.02&     0.1603&     0.0003&     2.91&     0.01&     2.13&     0.01\\
NGC104NS  &      1.71&     0.03&     3.49&     0.07&     0.1251&     0.0009&     2.55&     0.03&     2.02&     0.04\\
NGC362EW  &      1.83&     0.02&     2.62&     0.05&     0.0694&     0.0007&     0.94&     0.02&     1.34&     0.03\\
NGC362NS  &      1.90&     0.02&     2.59&     0.05&     0.0645&     0.0006&     1.03&     0.02&     1.36&     0.02\\
NGC3201EW &      2.06&     0.05&     2.11&     0.11&     0.0440&     0.0013&     1.01&     0.05&     0.99&     0.05\\
NGC3201NS &      2.85&     0.03&     1.28&     0.08&     0.0385&     0.0010&     0.93&     0.03&     0.84&     0.04\\
M68EW     &      2.59&     0.01&     0.81&     0.04&     0.0219&     0.0004&     0.64&     0.01&     0.58&     0.02\\
M68NS     &      2.26&     0.03&     0.63&     0.07&     0.0226&     0.0008&     0.61&     0.03&     0.41&     0.03\\
NGC4833EW &      2.13&     0.01&     1.41&     0.03&     0.0301&     0.0004&     0.46&     0.01&     0.79&     0.01\\
NGC4833NS &      2.50&     0.02&     1.11&     0.06&     0.0342&     0.0007&     0.38&     0.02&     0.58&     0.03\\
M5EW      &      2.50&     0.03&     2.15&     0.06&     0.0616&     0.0008&     1.13&     0.03&     1.30&     0.03\\
M5NS      &      2.80&     0.01&     1.94&     0.03&     0.0507&     0.0005&     1.08&     0.01&     1.14&     0.02\\
M80EW     &      2.24&     0.07&     1.57&     0.15&     0.0340&     0.0018&     0.84&     0.07&     0.88&     0.08\\
M80NS     &      2.37&     0.01&     1.47&     0.02&     0.0382&     0.0003&     0.87&     0.01&     0.92&     0.01\\
M4EW      &      1.88&     0.02&     2.87&     0.04&     0.1065&     0.0005&     2.10&     0.02&     1.60&     0.02\\
M4NS      &      2.86&     0.01&     2.99&     0.04&     0.0870&     0.0005&     1.46&     0.01&     1.60&     0.02\\
M12EW     &      2.23&     0.03&     1.75&     0.06&     0.0582&     0.0008&     1.35&     0.03&     1.06&     0.03\\
M12NS     &      2.93&     0.04&     1.34&     0.10&     0.0418&     0.0012&     1.07&     0.04&     0.83&     0.05\\
M10EW     &      2.52&     0.02&     1.45&     0.06&     0.0409&     0.0007&     0.86&     0.02&     0.84&     0.03\\
M10NS     &      2.96&     0.03&     1.50&     0.08&     0.0403&     0.0009&     0.78&     0.03&     0.82&     0.04\\
NGC6287EW &      2.59&     0.06&     0.34&     0.14&     0.0239&     0.0015&     0.47&     0.06&     0.30&     0.06\\
NGC6287NS &      3.14&     0.06&    -0.03&     0.12&     0.0244&     0.0013&     0.33&     0.05&     0.19&     0.05\\
NGC6293EW &      3.13&     0.03&     0.39&     0.08&     0.0220&     0.0009&     0.51&     0.03&     0.40&     0.04\\
NGC6293NS &      2.48&     0.02&     0.69&     0.06&     0.0268&     0.0007&     0.61&     0.02&     0.53&     0.03\\
NGC6342EW &      1.66&     0.07&     3.29&     0.15&     0.1295&     0.0018&     2.58&     0.06&     1.78&     0.07\\
NGC6342NS &      1.53&     0.04&     3.74&     0.09&     0.1334&     0.0011&     2.41&     0.04&     2.01&     0.04\\
NGC6352EW &      1.78&     0.03&     3.46&     0.06&     0.1134&     0.0008&     2.56&     0.03&     1.97&     0.03\\
NGC6352NS &      1.06&     0.03&     4.79&     0.06&     0.2417&     0.0008&     3.68&     0.02&     2.82&     0.02\\
NGC6362EW &      2.17&     0.02&     2.53&     0.06&     0.0619&     0.0008&     1.57&     0.03&     1.19&     0.03\\
NGC6362NS &      3.44&     0.02&     1.87&     0.04&     0.0488&     0.0005&     1.40&     0.02&     1.00&     0.02\\
NGC6397EW &      3.51&     0.06&     0.65&     0.14&     0.0210&     0.0016&     0.67&     0.06&     0.46&     0.07\\
NGC6397NS &      2.55&     0.03&     1.13&     0.06&     0.0273&     0.0007&     0.47&     0.03&     0.86&     0.03\\
NGC6528EW &      1.65&     0.07&     5.64&     0.14&     0.2375&     0.0018&     3.65&     0.06&     3.02&     0.07\\
NGC6528NS &      1.38&     0.07&     5.62&     0.13&     0.2266&     0.0016&     3.55&     0.06&     2.69&     0.06\\
NGC6541EW &      2.85&     0.04&     0.98&     0.10&     0.0319&     0.0012&     0.67&     0.04&     0.77&     0.05\\
NGC6541NS &      2.46&     0.02&     1.10&     0.05&     0.0339&     0.0006&     0.72&     0.02&     0.88&     0.02\\
NGC6553EW &      2.05&     0.04&     5.85&     0.07&     0.2368&     0.0009&     4.00&     0.03&     2.79&     0.03\\
NGC6553NS &      1.63&     0.04&     5.40&     0.08&     0.2086&     0.0010&     3.92&     0.03&     2.70&     0.03\\
M22EW     &      2.55&     0.01&     1.84&     0.02&     0.0688&     0.0003&     1.15&     0.01&     0.97&     0.01\\
M22NS     &      2.70&     0.04&     0.93&     0.09&     0.0272&     0.0010&     0.88&     0.04&     0.66&     0.04\\
M54EW     &      2.40&     0.04&     2.68&     0.10&     0.0625&     0.0012&     1.05&     0.04&     1.60&     0.05\\
M54NS     &      2.34&     0.02&     2.68&     0.05&     0.0636&     0.0006&     1.07&     0.02&     1.63&     0.02\\
NGC6752EW &      2.27&     0.01&     1.90&     0.04&     0.0445&     0.0005&     0.99&     0.02&     1.09&     0.02\\
NGC6752NS &      2.80&     0.02&     1.10&     0.05&     0.0373&     0.0007&     0.96&     0.02&     0.85&     0.03\\
M30EW     &      2.63&     0.04&     0.50&     0.09&     0.0176&     0.0011&     0.53&     0.04&     0.54&     0.04\\
M30NS     &      2.35&     0.02&     1.35&     0.04&     0.0487&     0.0005&     0.71&     0.02&     0.77&     0.02\\      
\hline
\label{table1}
\end{tabular}
The quoted error is the standard deviation from the propagation of the statistical error.
\end{flushleft}
\end{minipage}
\end{table*}


\begin{table*}
\centering
\begin{minipage}{120mm}
\begin{flushleft}
\caption[]{Uncorrected Lick indices averaged over the two directions}
\begin{tabular}{l|llllllllllll}
\hline
\hline
GC  &  H$_{\beta}  $&$\pm$&$  Fe5015  $&$\pm$&$  Mg_2  $&$\pm$&  Mg$_b$  &$\pm$&  Fe5270  &$\pm$\\
NGC104  &      1.66&     0.05&     3.65&     0.16&     0.142&     0.017&     2.73&     0.17&     2.07&     0.05   \\
NGC362  &      1.87&     0.03&     2.60&     0.01&     0.066&     0.002&     0.99&     0.04&     1.35&     0.00   \\
NGC3201  &     2.46&     0.39&     1.70&     0.41&     0.041&     0.002&     0.97&     0.03&     0.92&     0.07  \\
M68     &      2.43&     0.16&     0.72&     0.09&     0.022&     0.000&     0.62&     0.01&     0.50&     0.08      \\
NGC4833 &      2.31&     0.18&     1.26&     0.14&     0.032&     0.002&     0.42&     0.03&     0.68&     0.10  \\
M5     &       2.65&     0.14&     2.05&     0.10&     0.056&     0.005&     1.10&     0.02&     1.22&     0.08       \\
M80     &      2.31&     0.06&     1.52&     0.04&     0.036&     0.002&     0.85&     0.01&     0.90&     0.01      \\
M4      &      2.37&     0.48&     2.93&     0.05&     0.096&     0.009&     1.78&     0.32&     1.60&     0.00       \\
M12    &       2.58&     0.34&     1.55&     0.20&     0.050&     0.008&     1.21&     0.14&     0.94&     0.11      \\
M10     &      2.74&     0.22&     1.48&     0.02&     0.040&     0.000&     0.82&     0.03&     0.83&     0.01      \\
NGC6287 &      2.87&     0.27&     0.15&     0.19&     0.024&     0.000&     0.40&     0.07&     0.24&     0.05  \\
NGC6293 &      2.80&     0.32&     0.54&     0.14&     0.024&     0.002&     0.56&     0.05&     0.46&     0.06  \\
NGC6342 &      1.60&     0.06&     3.51&     0.22&     0.131&     0.001&     2.49&     0.08&     1.90&     0.11  \\
NGC6352 &      1.42&     0.35&     4.12&     0.66&     0.177&     0.064&     3.12&     0.56&     2.40&     0.42  \\
NGC6362 &      2.81&     0.63&     2.20&     0.33&     0.055&     0.006&     1.49&     0.08&     1.09&     0.09  \\
NGC6397 &      3.03&     0.48&     0.89&     0.24&     0.024&     0.003&     0.57&     0.09&     0.66&     0.19  \\
NGC6528 &      1.51&     0.13&     5.63&     0.00&     0.232&     0.005&     3.60&     0.05&     2.85&     0.16  \\
NGC6541 &      2.65&     0.19&     1.04&     0.05&     0.032&     0.001&     0.70&     0.02&     0.83&     0.05  \\
NGC6553 &      1.84&     0.20&     5.63&     0.22&     0.222&     0.014&     3.96&     0.04&     2.75&     0.04  \\
M22    &       2.63&     0.07&     1.38&     0.45&     0.048&     0.020&     1.01&     0.13&     0.81&     0.15      \\
M54  &         2.37&     0.02&     2.68&     0.00&     0.063&     0.000&     1.06&     0.01&     1.61&     0.01      \\
NGC6752  &     2.53&     0.26&     1.50&     0.40&     0.040&     0.003&     0.98&     0.01&     0.97&     0.11  \\
M30  &         2.49&     0.14&     0.92&     0.42&     0.033&     0.015&     0.62&     0.09&     0.66&     0.11\\   
\hline

\label{table2}
\end{tabular}
The quoted error is 1/2 of the difference in the index values between the two slit directions for each cluster.
\end{flushleft}
\end{minipage}
\end{table*}

\section{Effect of the correction to the Lick resolution}

In this Appendix we show the uncorrected indices averaged over the two directions measured as if our observational set-up were in the native Lick
resolution. We recall that, for ``uncorrected'' we mean data that have not (yet) been corrected for any offset between our and the Lick
system due to, e.g., residuals in the sky subtraction, systematics in the wavelength calibration. Therefore,
the entries in Table~\ref{table6} should be compared to those in Table~\ref{table2}.
Here we also measure the indices assuming that our resolution exactly matches the one of the Lick system (e.g. Worthey \& Ottaviani, 1997).
In fact, our resolution is slightly lower, therefore a small correction is in principle needed (see Sec. 3), however
we do not apply it here. The difference (Table~\ref{table6bis}) in the end-products is remarkably small, being
typically $<$1\% for H$_{\beta}$ and $Mg_2$, and 3-4\% for the other indices shown in the table.

\begin{table*}
\centering
\begin{minipage}{120mm}
\begin{flushleft}
\caption[]{Uncorrected indices averaged over the two directions -- Indices measured as if they were in the native Lick
resolution}
\begin{tabular}{l|llllllllllll}
\hline
\hline
GC  &  H$_{\beta}  $&$\pm$&$  Fe5015  $&$\pm$&$  Mg_2  $&$\pm$&  Mg$_b$  &$\pm$&  Fe5270  &$\pm$\\
NGC104   &      1.64&     0.05&     3.49&     0.15&     0.142&     0.017&     2.64&     0.17&     1.98&     0.05\\
NGC362   &      1.85&     0.03&     2.51&     0.01&     0.066&     0.002&     0.94&     0.04&     1.29&     0.00\\
NGC3201  &      2.43&     0.38&     1.66&     0.38&     0.040&     0.002&     0.92&     0.03&     0.87&     0.07\\
M68      &      2.40&     0.16&     0.75&     0.08&     0.022&     0.000&     0.59&     0.01&     0.47&     0.07\\
NGC4833  &      2.29&     0.18&     1.26&     0.13&     0.031&     0.002&     0.39&     0.03&     0.65&     0.10\\
M5       &      2.62&     0.14&     1.99&     0.09&     0.055&     0.005&     1.06&     0.02&     1.17&     0.07\\
M80      &      2.28&     0.06&     1.50&     0.04&     0.035&     0.002&     0.82&     0.01&     0.85&     0.01\\
M4       &      2.34&     0.47&     2.82&     0.05&     0.096&     0.009&     1.72&     0.31&     1.52&     0.00\\
M12      &      2.55&     0.34&     1.52&     0.19&     0.049&     0.008&     1.17&     0.13&     0.90&     0.11\\
M10      &      2.71&     0.21&     1.46&     0.02&     0.040&     0.000&     0.79&     0.03&     0.79&     0.01\\
NGC6287  &      2.83&     0.26&     0.21&     0.17&     0.023&     0.000&     0.38&     0.06&     0.23&     0.05\\
NGC6293  &      2.77&     0.32&     0.58&     0.13&     0.024&     0.002&     0.53&     0.05&     0.44&     0.06\\
NGC6342  &      1.58&     0.06&     3.36&     0.21&     0.130&     0.001&     2.41&     0.08&     1.81&     0.10\\
NGC6352  &      1.41&     0.35&     3.93&     0.62&     0.176&     0.064&     3.02&     0.54&     2.29&     0.40\\
NGC6362  &      2.77&     0.62&     2.14&     0.30&     0.055&     0.006&     1.43&     0.08&     1.04&     0.08\\
NGC6397  &      2.99&     0.47&     0.91&     0.22&     0.023&     0.003&     0.54&     0.09&     0.62&     0.18\\
NGC6528  &      1.50&     0.13&     5.34&     0.00&     0.231&     0.005&     3.48&     0.05&     2.73&     0.15\\
NGC6541  &      2.62&     0.18&     1.05&     0.05&     0.032&     0.001&     0.66&     0.02&     0.79&     0.05\\
NGC6553  &      1.83&     0.20&     5.34&     0.20&     0.222&     0.014&     3.83&     0.03&     2.62&     0.04\\
M22      &      2.60&     0.07&     1.37&     0.42&     0.047&     0.020&     0.97&     0.13&     0.77&     0.14\\
M54      &      2.35&     0.02&     2.58&     0.00&     0.062&     0.000&     1.02&     0.01&     1.54&     0.01\\
NGC6752  &      2.50&     0.25&     1.48&     0.37&     0.040&     0.003&     0.94&     0.01&     0.92&     0.11\\
M30      &      2.46&     0.14&     0.94&     0.39&     0.032&     0.015&     0.59&     0.08&     0.63&     0.11\\
\label{table6}
\end{tabular}
\end{flushleft}
\end{minipage}
\end{table*}

\begin{table*}
\centering
\begin{minipage}{120mm}
\begin{flushleft}
\caption[]{Applied sigma-corrections}
\begin{tabular}{l|llllllllllll}
\hline
\hline
GC  &  H$_{\beta}  $&  Fe5015  &$  Mg_2  $&  Mg$_b$  &  Fe5270  \\
NGC104   &      0.02&     0.16&     0.000&     0.09&     0.09\\
NGC362   &      0.02&     0.09&     0.000&     0.05&     0.06\\
NGC3201  &      0.03&     0.04&     0.001&     0.05&     0.05\\
M68      &      0.03&     0.03&     0.000&     0.03&     0.03\\
NGC4833  &      0.02&     0.00&     0.001&     0.03&     0.03\\
M5       &      0.03&     0.06&     0.001&     0.04&     0.05\\
M80      &      0.03&     0.02&     0.001&     0.03&     0.05\\
M4       &      0.03&     0.11&     0.000&     0.06&     0.08\\
M12      &      0.03&     0.03&     0.001&     0.04&     0.04\\
M10      &      0.03&     0.02&     0.000&     0.03&     0.04\\
NGC6287  &      0.04&     0.06&     0.001&     0.02&     0.01\\
NGC6293  &      0.03&     0.04&     0.000&     0.03&     0.02\\
NGC6342  &      0.02&     0.15&     0.001&     0.08&     0.09\\
NGC6352  &      0.01&     0.19&     0.001&     0.10&     0.11\\
NGC6362  &      0.04&     0.06&     0.000&     0.06&     0.05\\
NGC6397  &      0.04&     0.02&     0.001&     0.03&     0.04\\
NGC6528  &      0.01&     0.29&     0.001&     0.12&     0.12\\
NGC6541  &      0.03&     0.01&     0.000&     0.04&     0.04\\
NGC6553  &      0.01&     0.29&     0.000&     0.13&     0.13\\
M22      &      0.03&     0.01&     0.001&     0.04&     0.04\\
M54      &      0.02&     0.10&     0.001&     0.04&     0.07\\
NGC6752  &      0.03&     0.02&     0.000&     0.04&     0.05\\
M30      &      0.03&     0.02&     0.001&     0.03&     0.03\\
\label{table6bis}
\end{tabular}
\end{flushleft}
\end{minipage}
\end{table*}

\end{appendix}


\begin{thebibliography}{}

\bibitem[]{} Burstein D., Faber S.~M., Gaskell C.~M., 
Krumm N., 1984, ApJ, 287, 586 
\bibitem[]{} Cardiel N., Gorgas J., Cenarro J., Gonzalez J.~J., 1998, A\&AS, 127, 597 
\bibitem[]{} Carollo C.~M., Danziger I.~J., Buson L., 1993, MNRAS, 265, 553 
\bibitem[]{} Carretta E., Bragaglia A., Gratton R., 
Recio-Blanco A., Lucatello S., D'Orazi V., Cassisi S., 2010, arXiv, 
arXiv:1003.1723 
\bibitem[]{} Carretta E., Bragaglia A., Gratton R., 
Lucatello S., Bellazzini M., D'Orazi V., 2010, ApJ, 712, L21 
\bibitem[]{} Carretta E., Bragaglia A., Gratton R., D'Orazi V., Lucatello S., 2009, A\&A, 508, 695 
\bibitem[]{} Cohen J.~G., Blakeslee J.~P., Ryzhov A., 1998, ApJ, 496, 808 
\bibitem[]{} Covino S., Galletti S., Pasinetti L.~E., 1995, A\&A, 303, 79 
\bibitem[]{} Decressin T., Charbonnel C., Siess L., Palacios A., Meynet G., Georgy C., 2009, A\&A, 505, 727 
\bibitem[]{}Dekker, H., Delabre, B., D'Odorico, S., 1986, SPIE, 627, 39
\bibitem[]{} D'Ercole A., Vesperini E., D'Antona F., 
McMillan S.~L.~W., Recchi S., 2008, MNRAS, 391, 825 
\bibitem[]{} Gratton R., Sneden C., Carretta E., 2004, ARA\&A, 42, 385 
\bibitem[]{} Graves G.~J., Schiavon R.~P., 2008, ApJS, 177, 446 
\bibitem[]{} Greggio L., 1997, MNRAS, 285, 151 
\bibitem[]{}Harris, W.E. 1996, AJ, 112, 1487
\bibitem[]{} Lee H.-c., Worthey G., 2005, ApJS, 160, 176 
\bibitem[]{} Lee H.-c., Worthey G., Dotter A., 2009, AJ, 138, 1442 
\bibitem[]{} Lee H.-c., et al., 2009, ApJ, 694, 902 
\bibitem[]{} Lee J.-W., Kang Y.-W., Lee J., Lee Y.-W., 2009b, Natur, 462, 480 
\bibitem[]{} Maraston C., Greggio L., Renzini A., Ortolani S., Saglia R.~P., Puzia T.~H., Kissler-Patig M., 2003, A\&A, 400, 823 
\bibitem[]{} Matteucci F., 1994, A\&A, 288, 57 
\bibitem[]{} Matteucci F., 2001, ASSL, 253
\bibitem[]{} McWilliam A., Bernstein R.~A., 2008, ApJ, 684, 326 
\bibitem[]{} Mendel J.~T., Proctor R.~N., Forbes D.~A., 2007, MNRAS, 379, 1618
\bibitem[]{} Pipino A., Matteucci F., 2004, MNRAS, 347, 968
\bibitem[]{} Poole V., Worthey G., Lee H.-c., Serven J., 2010, AJ, 139, 809 
\bibitem[]{} Pritzl B.~J., Venn K.~A., Irwin M., 2005, AJ, 130, 2140 
\bibitem[]{} Puzia T.~H., Saglia R.~P., Kissler-Patig M., Maraston C., Greggio L., Renzini A., Ortolani S., 2002, A\&A, 395, 45\bibitem[]{}Renzini, A. 2007, ASPC, 380, 309
\bibitem]{}Salpeter E.~E., 1955, ApJ, 121, 161 
\bibitem[]{}Schiavon R.~P., 2007, ApJS, 171, 146 
\bibitem[]{} Schiavon R.~P., Rose J.~A., Courteau S., 
MacArthur L.~A., 2005, ApJS, 160, 163 
\bibitem[]{} Schiavon R.~P., Rose J.~A., Courteau S., 
MacArthur L.~A., 2004, ApJ, 608, L33 
\bibitem[]{} Serven J., Worthey G., 2010, AJ, 140, 152 
\bibitem[]{} Serven J., Worthey G., Briley M.~M., 2005, ApJ, 627, 754 
\bibitem[]{} Thomas D., Johansson J., Maraston C., 2010, arXiv, arXiv:1010.4570 
\bibitem[]{} Thomas D., Maraston C., Bender R., 2003, MNRAS, 339, 897 
\bibitem[]{} Thomas D., Maraston C., Bender R., Mendes de Oliveira C., 2005, ApJ, 621, 
673 
\bibitem[]{} Trager S.~C., Worthey G., Faber S.~M., Burstein D., Gonzalez J.~J., 1998, 
ApJS, 116, 1 
\bibitem[]{} Tripicco M.~J., Bell R.~A., 1995, AJ, 110, 3035 
\bibitem[]{} Ventura P., D'Antona F., 2009, A\&A, 499, 835 
\bibitem[]{} Worthey G., Faber S.~M., Gonzalez J.~J., 1992, ApJ, 398, 69 
\bibitem[]{} Worthey G., Faber S.~M., Gonzalez J.~J., 
Burstein D., 1994, ApJS, 94, 687 
\bibitem[]{} Worthey G., Ottaviani D.~L., 1997, ApJS, 111, 377 
\bibitem[]{} Zinn R., West M.~J., 1984, ApJS, 55, 45 

\end{thebibliography}
\end{document}